\documentclass[%
 reprint,
 superscriptaddress,
 amsmath,amssymb,
 aps,
]{revtex4-2}

\usepackage[dvipdfmx]{graphicx}
\graphicspath{{fig}}

\usepackage{dcolumn}
\usepackage{bm}

\usepackage{color}
\usepackage{soul}

\newcommand{\PdCaAgP}{$\mathrm{CaAg_{0.9}Pd_{0.1}P}$\ }

\newcommand{\sigmaN}{\sigma_\mathrm{N}}
\newcommand{\sigmaS}{\sigma_\mathrm{S}}
\newcommand{\sigmaT}{\sigma_\mathrm{T}}
\newcommand{\sigmaTC}{\sigma_\mathrm{T}^\mathrm{C}}

\newcommand{\Deltapx}{\mathrm{cos}\theta}
\newcommand{\Deltapy}{~\mathrm{sin}\theta}

\begin{document}

\title{Broken time-reversal symmetry detected by tunneling spectroscopy of superconducting Pd-doped CaAgP}

\author{Naoki Matsubara}
\email{matsubara.naoki.s5@s.mail.nagoya-u.ac.jp}
\affiliation{Department of Applied Physics, Nagoya University, Furo-cho, Chikusa, Nagoya, 464-8603 Japan}

\author{Rikizo Yano}
\affiliation{Department of Applied Physics, Nagoya University, Furo-cho, Chikusa, Nagoya, 464-8603 Japan}

\author{Kazushige Saigusa}
\affiliation{Department of Applied Physics, Nagoya University, Furo-cho, Chikusa, Nagoya, 464-8603 Japan}

\author{Koshi Takenaka}
\affiliation{Department of Applied Physics, Nagoya University, Furo-cho, Chikusa, Nagoya, 464-8603 Japan}

\author{Yoshihiko Okamoto}
\affiliation{Institute for Solid State Physics, University of Tokyo, Kashiwanoha 5-1-5, Kashiwa, Chiba, 277-8581 Japan}

\author{Yukio Tanaka}
\affiliation{Department of Applied Physics, Nagoya University, Furo-cho, Chikusa, Nagoya, 464-8603 Japan}

\author{Satoshi Kashiwaya}
\email{s.kashiwaya@nagoya-u.jp}
\affiliation{Department of Applied Physics, Nagoya University, Furo-cho, Chikusa, Nagoya, 464-8603 Japan}

\date{\today}

\begin{abstract}
The appearance of broken time-reversal symmetry (TRS) in superconducting states is an intriguing issue in solid-state physics
because of the incompatibility of the spontaneous magnetic field and the Meissner effect.
We identify broken TRS in Pd-doped CaAgP (\PdCaAgP) by tunneling spectroscopy through the magnetic field response of conductance spectra.
\PdCaAgP is a nodal-line semimetal with exotic electronic states such as drumhead surface states and surface superconductivity.
Tunneling conductance spectra acquired at the side surfaces of \PdCaAgP under an applied magnetic field exhibit broad zero-bias peaks with small asymmetric structures.
Surprisingly, the asymmetric structures are reversed exactly by flipping the field direction.
On the basis of an analysis 
which stands on the formula of tunneling junctions for unconventional superconductors,
these results are consistent with the pair potential of the superconductivity breaks the TRS and is strongly coupled to an external magnetic field.
We reveal the novel character of superconducting nodal-line semimetals by developing the TRS sensitivity of tunneling spectroscopy.
Our results serve as an exploration of broken TRS in superconducting states realized in topological materials.
\end{abstract}

\maketitle


\section{\label{sec:Intro}Introduction}
Symmetry breaking in nature is strongly related to the emergence of novel exotic states.
The onset of superconductivity due to the macroscopic coherent states is accompanied by the broken gauge symmetry, as described by Bardeen–Cooper–Schrieffer (BCS) theory~\cite{BCS}.
The onset of unconventional superconductivity is accompanied by extra symmetry breaking in momentum space and/or spin space, such as $d$-wave symmetry in high-$T_\mathrm{c}$ cuprates.
The exotic parity mixing due to antisymmetric spin--orbit coupling in noncentrosymmetric superconductors has recently been discussed~\cite{
PhysRevLett.97.017006,
PhysRevLett.92.097001,
PhysRevLett.87.037004}.
In the case of time-reversal symmetry (TRS) breaking of the macroscopic coherent states, the A-phase of superfluid $^3$He has been well established to break the TRS as a result of the chirality of the orbital motion~\cite{He3}.
However, it is unclear whether similar broken TRS states appear in superconductivity because the spontaneous magnetic field caused by the spontaneous current is incompatible with the Meissner effect.
Various candidates for the broken TRS in bulk superconducting Sr$_2$RuO$_4$~\cite{
	PhysRevLett.107.077003,
	PhysRevB.100.094530,
	Grinenko2021,
	PhysRevB.81.214501},
UPt$_3$~\cite{
	SchemmScience2014,
	PhysRevLett.71.1466,
	PhysRevLett.62.1411,
	StrandScience2010},
LaNiC$_2$~\cite{
	PhysRevLett.102.117007,
	Chen_2013},
UTe$_2$~\cite{
	UTe2.Nat.Commun,
	PhysRevB.100.140502,
	RanScience2019,
	HayesScience2021},
and Dirac semimetallic silicides~\cite{Silicide} are still under debate.
In this sense, broken TRS states localized near the surface, which do not conflict with the Meissner effect, are more feasible to be realized~\cite{JPSJ.64.3384, PhysRevX.4.031053}. 
\par
We focus on the topological nodal-line semimetal CaAgP, which is a potential candidate for an unconventional superconductor
~\cite{PhysRevB.102.115101,Yano2023}.
In fact, several exotic characteristics of nodal-line semimetals have been experimentally detected
~\cite{Aggarwal_2019, HiraiJPSJ2022}.
CaAgP has a hexagonal noncentrosymmetric crystal structure belonging to $P\bar{6}2m$.
This material is ideal for the investigation of nodal-line semimetals because the Fermi level ($E_\mathrm{F}$) is located near the 
nodal ring
and no additional bulk bands other than the nodal Dirac band exist near the $E_\mathrm{F}$~\cite{YamakageJPSJ2016, Salmankurt2023}.
As a result of the fine-tuning of the $E_\mathrm{F}$ by Pd doping (Pd-doped CaAgP, referred to here as \PdCaAgP), the $E_\mathrm{F}$is estimated to be 0.14 eV below the nodal ring. 
According to theoretical calculations, \PdCaAgP has a weakly dispersive drumhead surface band in addition to the bulk torus-shaped Fermi surface (Fig. 1(a))~\cite{YamakageJPSJ2016}.
In fact, its transport properties are consistent with the two-carrier model of dominant hole carriers due to the bulk band and extremely high-mobility electron carriers due to the surface band.
The superconductivity of \PdCaAgP (critical temperature $T_\mathrm{c} \sim 1.8$~K) has been suggested to be unconventional and localized at the surface on the basis of heat capacity measurements and the quantum oscillation presented in previous works~\cite{PhysRevB.102.115101,Yano2023}.
Tunneling spectroscopy based on the soft point-contact method shows dome-shaped conductance peaks indicating topological superconductivity similar to that observed in Sr$_2$RuO$_4$~\cite{PhysRevLett.107.077003} and Cu-doped Bi$_2$Se$_3$~\cite{PhysRevLett.107.217001}.
\par
In this paper, we further investigate the tunneling spectroscopy of  \PdCaAgP using normal metal/insulator/superconductor (N/I/S) junctions to clarify the detailed superconducting properties.
The focus is on the asymmetries in the bias voltage ($V$) dependence of the differential conductance spectra and their magnetic field ($B$) responses detected in experiments.
We find that flipping the magnetic field direction exactly reverses the small asymmetric structures in experimental conductance spectra of positive and negative bias.
To analyze this feature, researchers have applied the extended Blonder--Tinkham--Klapwijk(BTK) formula to include the effects of broken TRS of the pair potential and asymmetric tunneling current distribution~\cite{BTK, PhysRevLett.74.3451, Kashiwaya1996,PhysRevB.56.7847, Kashiwaya2000}.
The experimentally detected responses are consistent with theoretical formulas for the pair potential of broken TRS coupling to the external field. 
These results clarify the exotic symmetry breaking of the superconducting nodal-line semimetals and demonstrate the novel ability to detect the broken TRS of the pair potential by tunneling spectroscopy.
\par

\section{Tunneling spectroscopy experiments of \PdCaAgP}
Single crystals of \PdCaAgP were grown by a flux method and have hexagonal prism shapes extending along the $c$-axis~\cite{Yano2023}.
Because the surfaces were stable in air, we could easily fabricate N/I/S junctions using the soft point-contact technique~\cite{PhysRevLett.107.217001,PhysRevLett.89.247004}.
Among the junctions we fabricated, we refer to four: \#1-Top, \#1-Side, \#2-Side, and \#3-Side, where \#i-Top (Side) indicates a junction formed on the top (side) surface of crystal \#i ($1\le i\le 3$), as shown in Fig. 1(b).
Conductance spectra (d$I$/d$V$--$V$) were acquired using the usual four-terminal method with a lock-in amplifier (LI-5640, NF); the sample was maintained at temperatures as low as 0.6 K in a $^3$He refrigerator.
\par
Figures~1(c)--(f) show the conductance spectra for \PdCaAgP in the absence of an applied field.
The peaks in the conductance spectra appear at temperatures below $T_\mathrm{c}$, suggesting that their origin is the superconductivity.
The value of  $2\Delta _{\mathrm{G}} / \left( k_{\text{B}}T_{\text{c}} \right) \sim 5.16$ where $\Delta _{\mathrm{G}}$ is the gap amplitude evaluated on the basis of the distance between the two side dips is in the range reasonable for unconventional superconductivity.
All of the conductance spectra show broad zero-bias peaks at the lowest temperature, consistent with the previous result.
Although the features in the spectra at the top and side junctions appear similar, the magnetic field response differs, as described below.
The broad zero-bias peak, because of the appearance of the dispersive surface Andreev bound state, suggests topological superconductivity such as chiral $p$-wave or helical $p$-wave symmetries~\cite{PhysRevLett.107.077003,KASHIWAYA201425}.
Another possible origin for the peak is the heating effect; the conductance peaks appear as the tunneling current exceeds the critical current density of the superconductor when the junction is in a thermal regime~\cite{PhysRevB.69.134507}.
Such a mechanism can be rejected on the basis of the magnetic field dependence of the zero-bias peak heights and the estimation of the critical current density in the junction~\cite{Yano2023}. 
In addition, because of the heating effect, we cannot expect systematic magnetic field responses of the conductance spectra described below. 
\par
\begin{figure*}[htb]
\includegraphics[keepaspectratio,width=1.8\columnwidth]{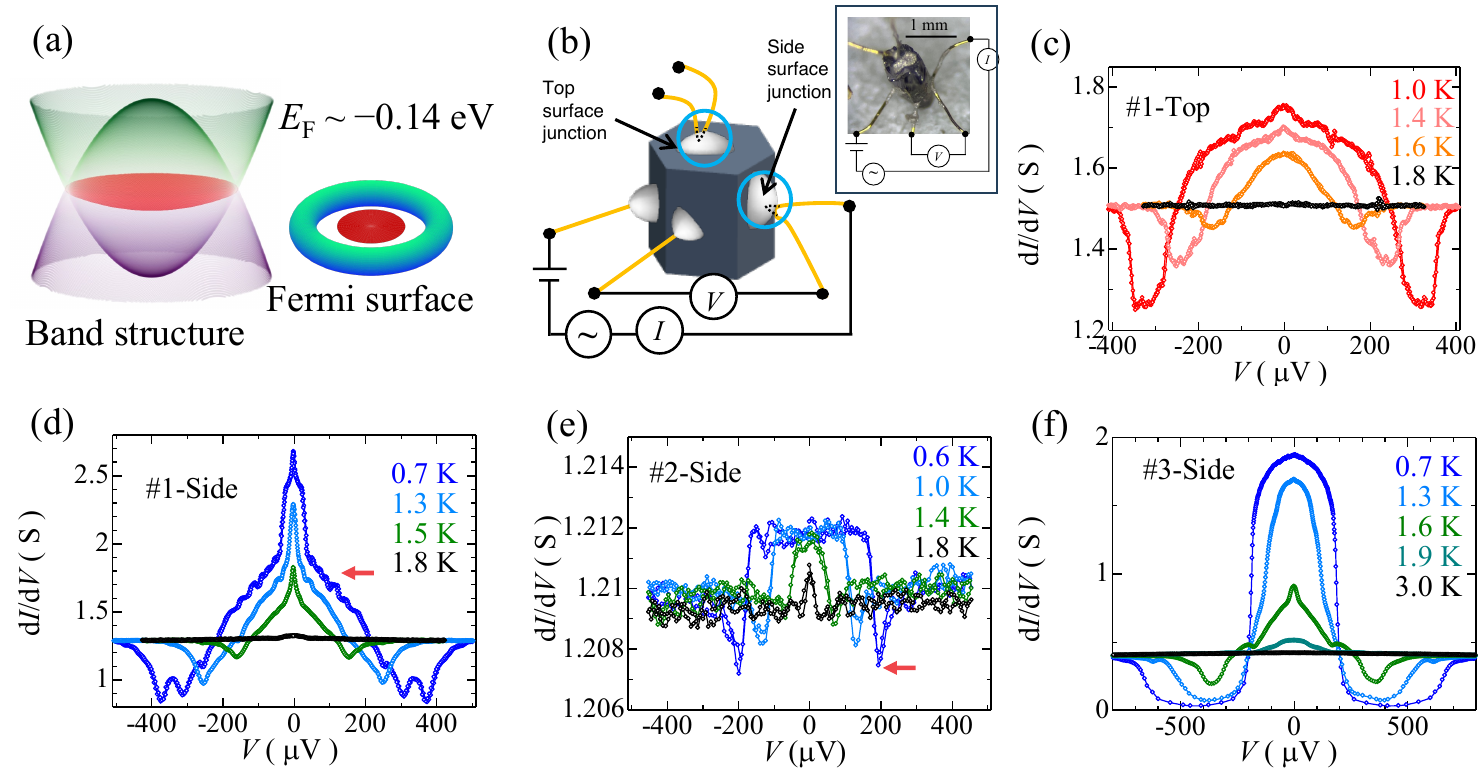}
	\caption{\label{fig1}%
		(a) Schematic of the band structure and Fermi surface of \PdCaAgP with $ E_\mathrm{F} \sim -0.14~\mathrm{eV}$ below the Dirac node.
		Red indicates the surface electron state, and green and purple show the bulk band.
		The Fermi surface is composed of the bulk toroid and the surface flat-dispersive plane.
		(b) Schematic of a \PdCaAgP crystal and the soft point contacts formed at the top and side surfaces.
		(c--f) The temperature dependences of conductance spectra obtained at four junctions (\#1-Top(c), \#1-Side(d), \#2-Side(e), and \#3-Side(f)) in the absence of an applied field.
		Here, \#i-Top (Side) indicates the junction formed on the top (side) surface of crystal \#i ($1\le i\le 3$).
		In all cases, broad zero-bias conductance peaks appear below $ T_\mathrm{c} \sim 1.8~\mathrm{K}$, indicating unconventional superconductivity of \PdCaAgP.
		}
\end{figure*}
We hereafter focus on the small structures in conductance spectra and their magnetic field responses.
The overall features of the zero-bias peaks are symmetric, whereas asymmetric small structures are detected in the side junctions.
They are the dip and hump in Fig. 1(d) and the dips at approximately $\pm200~\mathrm{\mu V}$ in Fig.~1(e) (indicated by red arrows).
We examine the magnetic field responses and their orientational dependencies to clarify the origin of these asymmetries.
Figure~2(a) shows the magnetic field response of conductance spectra corresponding to the N/I/S junction on \#1-Top.
At the top surface, although the spectra show small asymmetric structures, they tend to be smeared when a magnetic field is applied.
However, the responses at the side surfaces are completely different; 
the dip and hump structures in \#1-Side (Fig.~2(b)) and \#3-Side (Fig.~2(d)), indicated by red arrows, and  the two asymmetric dips at $\pm 200\mathrm{\mu V}$ in \#2-Side (Fig.~2(c)) appear exactly at the reversed bias side when the magnetic field direction is flipped.
To trace how this reversal occurs, we evaluated the asymmetric factor $\gamma (B)$ as the function of the applied field defined by
\begin{align}
\gamma (B)=\mathrm{d}I/\mathrm{d}V(+\Delta _\mathrm{G})-\mathrm{d}I/\mathrm{d}V(-\Delta_\mathrm{G}).
\end{align}
The factors $\gamma (B)$ shown in Figs.~2(e) and 2(f) indicate that the reversing occurs only near the zero field.
In addition, the switching-type transition between the two states shown in \#2-Side (Fig.~2(e)) differs completely from the linear-type responses.
These features are unique because the tunneling conductance spectra have been reported to preserve positive and negative bias voltage symmetry (\textit{i.e.}, electron--hole symmetry) for various pair potential symmetries, including those that break the TRS based on the conventional formula of tunneling conductance ~\cite{KASHIWAYA201425}.
In fact, several asymmetric spectra corresponding to N/I/S junctions of unconventional superconductors have been reported previously, including those of high-$T_\mathrm{c}$ superconductors~\cite{ANDERSON20061} and heavy fermion superconductors~\cite{W.K.Park_2009}.
However, reversal of the conductance spectra with the switching-type transitions has not been reported.
Thus, the origin of the asymmetric spectra acquired in the present study should differ from the origins of asymmetric spectra described in previous works.
\par
We here discuss possible origins of the asymmetry.
The dominant magnetic field response that appears in conductance spectra is the Doppler shift, in which the quasiparticles experience an energy shift due to the finite shielding current ~\cite{PhysRevLett.79.281}.
On the basis of the Doppler shift, response is expected to be the energy shift or the smearing of dip/hump structures linearly to the applied field.
This response is likely consistent with the response of \#1-Top but apparently inconsistent with those of the side junctions.
In addition, the typical response scale of several meV/T in the Doppler shift~\cite{PhysRevLett.79.281} is far smaller than that of the present experimental response.
Another possible origin is the nonreciprocal effect expected for systems with broken inversion symmetry systems ~\cite{Ideue2017,TokuraNagaosa2018}.
This effect is possible in the sense that the crystal structure of CaAgP lacks inversion symmetry.
However, the response should be linear to the field, which is inconsistent with the switching behavior in the present junctions.
Therefore, we consider that the most plausible origin is that, similar to chiral $p$-wave superconductivity, the superconductivity of \PdCaAgP breaks TRS and is strongly coupled to the external field.
The switching-type transition can be explained if the pair potential switches to a reversed chirality state by flipping the field.
Other broken TRS states, such as the $d \pm is$-wave state, are candidates. 
However, the spectral shapes shown in Fig.~2 are incompatible with the $d \pm is$-wave state.
Therefore, in the following discussion, chiral $p$-wave superconductivity is assumed to be the broken TRS state of 
$\mathrm{CaAg_{0.9}Pd_{0.1}P}$.
\begin{figure*}
\includegraphics[keepaspectratio,width=1.8\columnwidth]{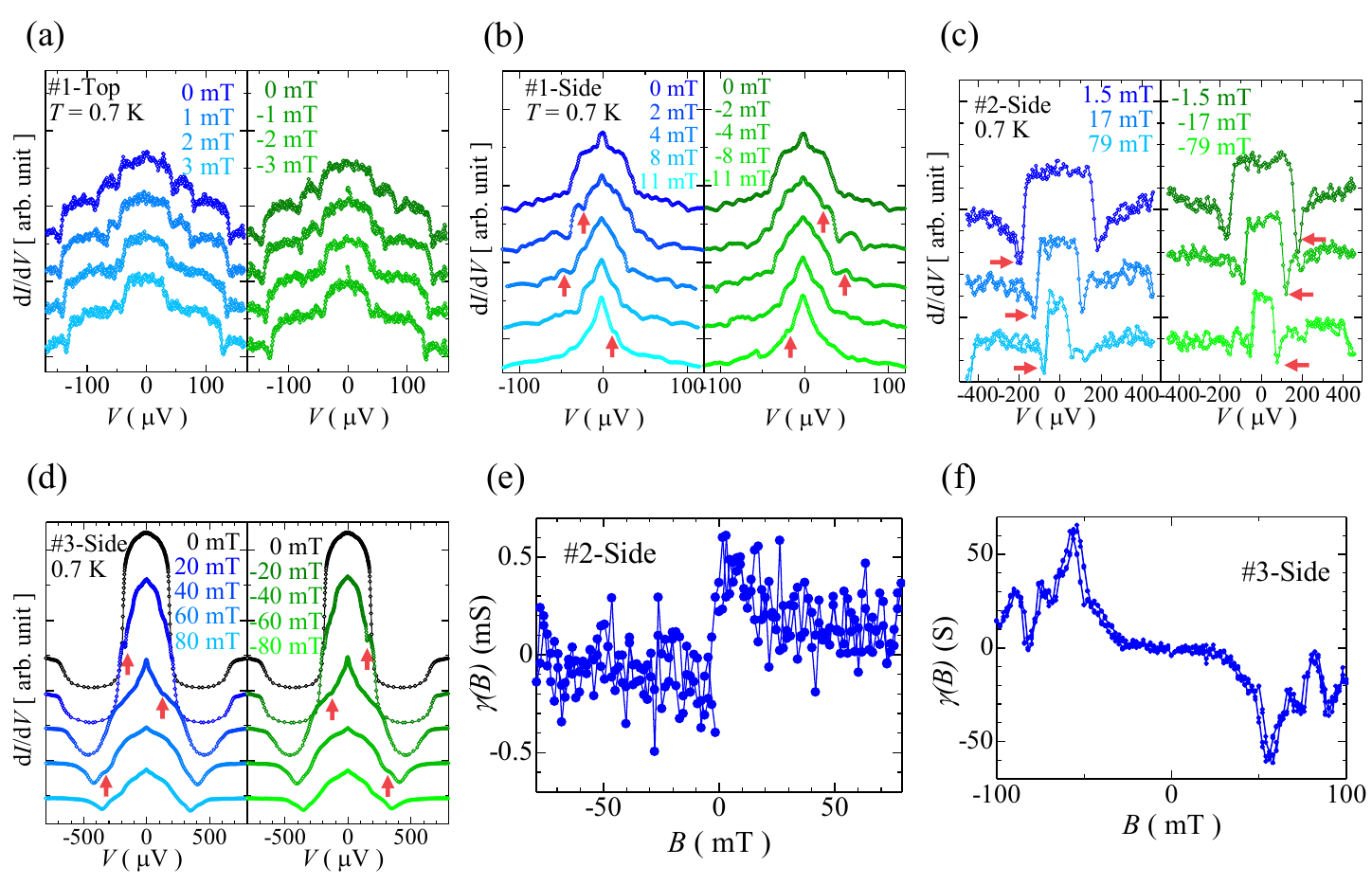}
	\caption{%
		(a--d) The magnetic field dependences of conductance spectra for 
		(a) \#1-Top, (b)\#1-Side, (c) \#2-Side, and (d) \#3-Side junctions.
		The fine structures in (a)\#1-Top are simply smeared by application of the field.
		However, asymmetric small structures that appear in side junctions exhibit peculiar responses.
		Flipping the field direction exactly reverses the structures of positive and negative bias, as indicated by the red arrows.
		(e,f) The asymmetric factor $\gamma (B)$ defined by equation (1) for (e) \#1-Side and (f) \#2-Side are plotted. 
		}
\end{figure*}

\section{Theoretical formula of conductance spectra for broken TRS superconductors}
In this section, we examine the conductance spectra of N/I/S junctions for unconventional superconductors formulated on the basis of the extended BTK formula ~\cite{BTK, PhysRevLett.74.3451, Kashiwaya1996,PhysRevB.56.7847,Kashiwaya2000,KASHIWAYA201425}.
Because the energy scale of the magnetic field is small compared with the gap amplitude, the effect of the vector potential is neglected.
The extended BTK formula assumes quasiparticle injection from the normal side, and the conductance is calculated by the normal reflection and Andreev reflection probabilities.
The electron injection process is treated in a two-dimensional plane by assuming translational invariance for the $y$ and $z$ directions (Fig.~3(a)).
For a quasiparticle injection with energy $E$ ($=-eV$, where $e$ is the elementary charge) and angle $\theta$ to the interface normal, the angle-resolved conductance $\sigmaS(E,\theta)$ is given by ~\cite{PhysRevLett.74.3451,Kashiwaya2000}
\begin{widetext}
\begin{align}
        \sigmaS(E,\theta) &= 
		\frac{1+\sigmaN(\theta)|\Gamma_{+}(E,\theta)|^2
				+(\sigmaN(\theta)-1)|\Gamma_{+}(E,\theta)\Gamma_{-}(E,\theta)|^2}
			{|1+(\sigmaN(\theta)-1)\Gamma_{+}(E,\theta)\Gamma_{-}(E,\theta)|^2},\\
        \Gamma_{+}(E,\theta)&=\frac{\Delta_{+}^{*}(\theta)}{E+\Omega_{+}(E,\theta)},
        \Gamma_{-}(E,\theta)=\frac{\Delta_{-}(\theta)}{E+\Omega_{-}(E,\theta)},\\
        \Omega_{\pm}(E,\theta)&=
        \begin{cases}
            \sqrt{E^2-|\Delta_\pm(\theta)|^2} & (E>|\Delta_\pm(\theta)|),\\
            i\sqrt{|\Delta_\pm(\theta)|^2-E^2} &( -|\Delta_\pm(\theta)|\leq E\leq |\Delta_\pm(\theta)|),\\
            -\sqrt{E^2-|\Delta_\pm(\theta)|^2} & (E<-|\Delta_\pm(\theta)|).
        \end{cases}
\end{align}
\end{widetext}
Here, $\sigmaN(\theta)$ and $\Delta_{+(-)}(\theta)$ are the conductance in a normal state and the effective pair potentials for the transmitted electron(hole)-like quasiparticles, respectively.
The total tunneling conductance $\sigmaT(E)$ is obtained by integration for all injection angles,
\begin{align}\label{eq:sigmaT}
\sigmaT(E) =
\frac{\int_{-\pi/2}^{\pi/2}\sigmaS(E,\theta)
\sigmaN(\theta)\cos\theta \mathrm{d}\theta}
{\int_{\pi/2}^{-\pi/2}\sigmaN(\theta)\cos\theta \mathrm{d}\theta}.
\end{align}
\par
The electron--hole symmetry in conductance spectra ($\sigmaT(E) =\sigmaT(-E)$) are analyzed on the basis of the properties of $\sigmaN(\theta)$ and $\sigmaS(E, \theta)$.
First, $\sigmaN(\theta)$ represents the distribution of tunneling quasiparticles, which is determined by the electronic structure of the electrodes and by a potential barrier independent of the pair potential.
In most cases, because the isotropic electronic states and perfectly translational symmetric $y$- and $z$-axes are assumed, $\sigmaN(\theta)$ is a symmetric function of the injection angle $\theta$ ($\sigmaN(\theta) = \sigmaN(-\theta)$).
In fact, for a delta-functional barrier $H_\mathrm{b} \delta(x)$, $\sigmaN(\theta) = \frac{\cos^2\theta}{Z^2+\cos^2\theta}$ (where $Z$, $m$, and $\hbar$ are $\frac{mH_\mathrm{b}}{\hbar^2 k_\mathrm{F}}$, the electron mass, and the Dirac constant, respectively) is symmetric to $\theta$.
This symmetry is evident because the junction geometry is $\theta$-symmetric.
By contrast, in real junctions, the anisotropy in electronic states and roughness at the interface break the geometric symmetry; thus, the experiments should be carried out under the condition $\sigmaN(\theta) \neq \sigmaN(-\theta)$ (refer to Appendix A for additional details).
Next, the properties of $\sigmaS(E, \theta)$ are dominated by the symmetry of the pair potential $\Delta(\theta)$.
When the pair potential is TRS invariant (TRSI, $\Delta(\theta)=\Delta^{*}(\theta)$),
$\sigmaS(E,\theta)=\sigmaS(-E,\theta)=\sigmaS(E,-\theta)=\sigmaS(-E,-\theta)$ is always satisfied.
However, when the pair potential is TRS-broken (TRSB, $\Delta(\theta)\neq\Delta^{*}(\theta)$),
$\sigmaS(E,\theta)=\sigmaS(-E,-\theta) \neq \sigmaS(E,-\theta)=\sigmaS(-E,\theta)$ is applied (refer to Appendix B for additional details).
\par
The electron--hole symmetries in conductance spectra, as evaluated under the above consideration, are summarized in Table~I.
For TRSI superconductors, the electron--hole symmetry is satisfied independent of $\sigmaN(\theta)$, whereas, for TRSB superconductors, the electron--hole symmetry is lost when $\sigmaN(\theta)$ is not $\theta$-symmetric.
In addition, the broken electron--hole symmetry in this mechanism can be discriminated from that of other origins by checking the relation $\sigmaT(E) =\sigmaTC (-E)$, where $\sigmaTC(E)$ corresponds to the conductance spectra for the time-reversed pair potential $\Delta^*(\theta)$.
Such measurements can be easily carried out in experiments by flipping the magnetic field if the pair potential is strongly coupled to the external field.
The properties of $\sigmaN(\theta)$ are insensitive to the magnetic field when the junction is nonmagnetic.
\par
\begin{table}[bt]
\centering
\caption{
	The electron--hole symmetries in $\sigmaT (E)$ are summarized for various cases. 
	In the table, TRSI and TRSB correspond to the cases of $\Delta(\theta) =\Delta^*(\theta)$ and $\Delta(\theta) \neq \Delta^*(\theta)$, respectively, and $\sigmaTC$ indicates the conductance for the time-reversed pair potential $\Delta^*(\theta)$.}
\label{tb:sigmaT}
\begin{tabular}{c||c|c}
	&Symmetric $\sigmaN (\theta) $
	&Asymmetric $\sigmaN (\theta)$ \\ \hline\hline
	TRSI
	&i) $\sigmaT (E) = \sigmaT(-E)$
	&iii) $\sigmaT (E) = \sigmaT(-E)$\\ \hline
	TRSB
	&ii) $\sigmaT (E) = \sigmaT(-E)$
	&\begin{tabular}{c} \rm{iv)}  $\sigmaT (E) \neq \sigmaT(-E)$ \\
	$ \sigmaT (E) = \sigmaTC(-E)$    \end{tabular}
\end{tabular}
\end{table}

\begin{figure*}
\includegraphics[keepaspectratio,width=1.8\columnwidth]{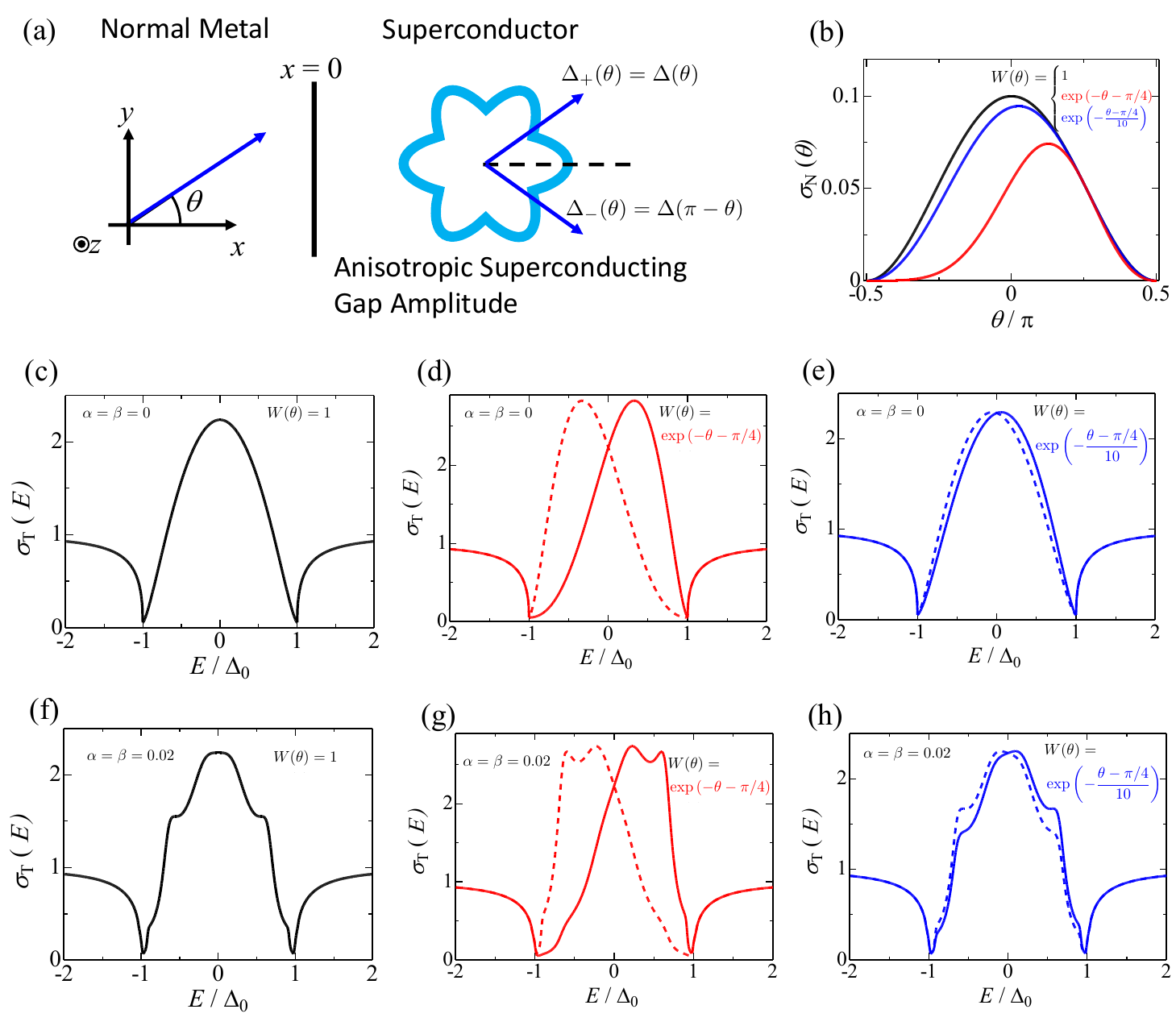}
	\caption{
		(a) Schematics of a normal metal/insulator/superconductor junction assumed in the extended BTK formula.
		The superconductor has an anisotropic gap amplitude, as illustrated on the right side.
		The blue arrows indicate the group velocity of injected quasiparticles and transmitted electron (hole)-like quasiparticles. 
		(b) 	Three types of tunneling current distribution $\sigmaN(\theta)$ used in the simulation.
		Compared with the symmetric $\sigmaN(\theta)$ for the $\delta$-function barrier model, blue and red curves contain the weighting factors described in equation (7).
		(c--h) Results of the simulation of the tunneling conductance $\sigmaT(E)$ with
		$\Delta(\theta) = \Delta_0 (1+\alpha\cos(6\theta)+\beta\cos(12\theta))(\Deltapx+i\Deltapy)$.
		(c--e) for $\alpha=\beta=0$,
		(f--h) for $\alpha=\beta=0.02$.
		The color of each plot corresponds to the color of $\sigmaN(\theta)$ shown in (b).
		The solid and dotted lines correspond to the results for $\Delta(\theta)$ and $\Delta^*(\theta)$, respectively.
		The exact reversing relation of the two curves reflects the relation $\sigmaT(E)=\sigmaTC(-E)$ of iv) in Table I.
		}
\end{figure*}
\par
We analyzed experimental results on the basis of a simulation of the extended BTK formula to clarify the correspondence between theory and experiments.
We consider chiral $p$-wave ($\Delta(\theta) = \Deltapx +i \Deltapy$) symmetry for the pair potential because the experimentally acquired spectra are consistent with the helical and chiral cases among various candidates ~\cite{KASHIWAYA201425}.
For the origin of the small structures in the experimental spectra, we introduce the anisotropy of the gap amplitude described by
$\Delta(\theta) = \Delta_0 (1+\alpha\cos(6\theta)+\beta\cos(12\theta))(\Deltapx+i\Deltapy)$
by considering the hexagonal crystal structure of CaAgP (Fig.~3(a)).
Figures~3(c) and (f) show the conductance spectra for various anisotropies of the gap amplitude with symmetric $\sigmaN(\theta)$ ($Z=3$).
Compared with the isotopic ($\alpha=\beta=0$) case shown in Fig.~3(c), small structures appear similar to that detected in \#1-Side (Fig.~2(b)) when a small anisotropy ($\alpha=\beta=0.02$) is introduced, as shown in Fig.~3(f), whereas the spectra preserve the electron--hole symmetry consistent with (ii) in Table I.
\par
Next, we consider the effect of asymmetric $\sigmaN(\theta)$.
Here, we phenomenologically treat the anisotopic effect to analyze qualitative features of the magnetic field responses shown in Fig.~2.
Because the exact determination of $\sigmaN(\theta)$ in real experimental situations is impossible, we employ asymmetric $\sigmaN(\theta) = \frac{\cos^2\theta}{Z^2+\cos^2\theta} \times W(\theta)$, where $W(\theta)$ is the weighting factor given by
\begin{equation}
W(\theta) =
\begin{cases}
	1 &(\mathrm{black}),\\
	\exp\left(-(\theta-\pi/4)\right) &(\mathrm{red}),\\
	\exp\left(-\frac{\theta-\pi/4}{10}\right) &(\mathrm{blue})\\
\end{cases}
\end{equation}
as described in Fig.~3(b).
The solid lines in Figs.~3(c)--(e) show the conductance spectra $\sigmaT(E)$ for the chiral $p$-wave pair potential ($\Delta(\theta)=\Delta_0(\Deltapx+i\Deltapy)$) calculated for three types of $\sigmaN(\theta)$.
The electron--hole symmetry in the conductance spectra is clearly broken by introduction of the asymmetry.
The asymmetric conductance spectra are just reversed by flipping the chirality $\Delta^*(\theta)$, as described by the dotted lines.
These features are clearer for the pair potential with the small anisotropy employed above.
The solid lines in Figs. 3(f)--(h) show the conductance spectra with small anisotropy calculated for various $\sigmaN(\theta)$.
The broken electron--hole symmetries are reproduced for asymmetric $\sigmaN(\theta)$.
The small structures are flipped by reversing the chirality, as shown by the dotted lines.
This exact reversing of the positive and negative energy in the conductance spectra is expressed by  $\sigmaT(E) = \sigmaTC(-E)$ and is consistent with iv) in Table I.
\par
Comparing the experimental results with the above simulation results, we find that the peculiar features experimentally detected are consistent with those for chiral $p$-wave superconductivity with asymmetric $\sigmaN(\theta)$.
On the basis of these considerations, we conclude that the pair potential of Pd-CaAgP breaks the TRS.
Note that the chirality flipping occurs near the zero-magnetic field (Fig.~2(e)).
We consider that the pair potential of Pd-CaAgP is strongly coupled to the external field because of its surface superconductivity.
We also note that a similar magnetic field response should appear in the broken TRS superconductor even if the pair potential is not chiral $p$-wave symmetry.
\par

\section{Discussion}
Asymmetry in tunneling conductance spectra has been previously reported in the spectra of UTe$_2$ acquired by scanning tunneling spectroscopy~\cite{UTe2.Nat.Commun}.
We consider that the mechanism is similar to our situation; however, in the previous report, the asymmetry of the spectra was caused by the step structure of the superconductor; in addition, switching of the chirality by an external field was not detected.
In another related report, Chiu $et$ $al.$ detected the broken TRS superconductivity as a T-shaped junction of CoS$_2$/TiS$_2$~\cite{ChiuSciAdv2021}.
They reported the appearance of a hysteresis loop in the conductance spectrum when an external field was applied, which is different from the results of the present study.
In fact, how chirality flipping occurs in a magnetic field depends on the condensation energy of superconductivity and the strength of the coupling; thus, it is dependent on the shape of the superconductor and the symmetry of the pair potential.
Therefore, we need to examine this issue further.
\par
Finally, we discuss the origin of the TRS breaking in superconducting \PdCaAgP.
The broken TRS is less likely to be caused by non-unitary superconductivity because \PdCaAgP is not a magnetic material.
Therefore, the multicomponent order parameter is reasonable for the origin of broken TRS superconductivity.
Because \PdCaAgP has a non-centrosymmetric crystal structure, the mixing of singlet and triplet order due to the antisymmetric spin--orbit interaction called ``anapole superconductivity'' is a possible scenario ~\cite{KanasugiYanase2022}.
Another scenario is that the coexisting bulk and surface drumhead bands induce novel superconductivity at the surface.
The drumhead surface states effectively correspond to a flat dispersive band, which is a potential platform for the ferromagnet and high-$T_\mathrm{c}$ superconducting order~\cite{PhysRevX.11.031017}.
The mixing of the bulk toroidal states with the flat band can induce novel electronic states.
Because systematic Pd doping can tune the Fermi level and, thus, the band mixing, it will provide important clues regarding the mechanism of superconductivity.
\par

\section{Conclusion}
We carried out soft point-contact spectroscopy of the superconducting nodal-line semimetal \PdCaAgP.
The conductance spectra show broad zero-bias peaks due to the topological superconductivity of this material.
Tunneling conductance spectra acquired at the \PdCaAgP surfaces parallel to the $c$-axis under an applied magnetic field exhibit asymmetric fine structures, indicating broken electron--hole symmetry.
The asymmetric structure of the positive and negative bias voltage is exactly reversed by flipping the field direction.
We applied the extended BTK formula to analyze these features to include the broken TRS and asymmetric tunneling current distribution. 
The anomalous reversal of the fine structures in the spectra is consistent with the reversal of the chirality through flipping of the applied magnetic field.
These results verify the broken TRS in superconducting \PdCaAgP and demonstrate the novel ability to identify the TRS using tunneling spectroscopy.
A comparison with the other direct probes for broken TRS, such as $\mu$SR measurements and the Kerr effect, are indispensable for further establishing the present results.
Also, systematic doping of Pd into CaAgP to tune the Fermi level can reveal the novel properties of the surface drumhead bands mixing with the bulk toroidal states for the onset of superconductivity.

\appendix

\section{A: Asymmetry in tunneling electron density distribution}

This section examines the properties of the tunneling current distribution in a tunneling junction described by $\sigmaN(\theta)$.
As a model, we consider an N$_\mathrm{L}$/I/N$_\mathrm{R}$ junction in which both electrodes are in their respective normal state.
Conventional tunneling models treat two-dimensional $xy$-planes with translational symmetry in the $y$-axis and the barrier potentials described by a delta-functional form $H_\mathrm{b} \delta(x)$ (Fig.~4(a)).
Isotropic circular Fermi surfaces are assumed for both electrodes, although the sizes of the Fermi surfaces in the two electrodes can differ ~\cite{PhysRevLett.74.3451, Kashiwaya1996, Kashiwaya2000, Linder2008, Cayssol2008, Breunig2021}.
The tunneling current distribution $\sigmaN(\theta)$ is given by the transmission probability amplitude for the quasiparticle injection from the left side with an angle $\theta$ ($-\pi/2<\theta<\pi/2$) to the interface normal.
Because the energy of the quasiparticle is assumed to be almost equivalent to the Fermi energy ($E \sim E_\mathrm{F}$), the energy dependence can be neglected.
In this model, because the system is $\theta$-inversion symmetric and $\sigmaN(\theta)$ is a real positive value, $\sigmaN(\theta)$ also has $\theta$-inversion symmetry ($\sigmaN(\theta)=\sigmaN(-\theta)$).
When the sizes of the Fermi surfaces of both electrodes are equal, $\sigmaN(\theta)$ is given by
\begin{align}
        \sigmaN(\theta) = \frac{\cos^2\theta}{Z^2+\cos^2\theta},
\end{align}
where $Z = mH_\mathrm{b}/\hbar^2k_\mathrm{F}$ ($m$, $\hbar$, and $k_\mathrm{F}$ are the electron mass, Dirac constant, and Fermi wavelength, respectively).
\par
Next, we consider the case in which the system breaks the $\theta$-inversion symmetry.
One example is the junction with a non-flat interface.
$\sigmaN(\theta)$ clearly loses the $\theta$-inversion symmetry.
Unfortunately, we are unable to analytically derive the tunneling current distribution.
Another example is the Fermi surface having a non-circular shape, as described in Fig.~4(b).
Because the system has translational invariance for the $y$ (and $z$) direction, the momenta along the $y$ and $z$ axes are conserved on the basis of Neother's theorem.
We assume that the Fermi energies are equal.
The conservation of the $k_y$ component is described by
\begin{align}
        k_\mathrm{R}\sin\theta=k_\mathrm{L}\sin \phi,
\end{align}
where $k_\mathrm{L(R)}$ is the wavelength of the left (right) electrode.
The wave functions of the left and right electrodes are given by
\begin{equation}
\begin{cases}
	\varphi_\mathrm{L}(x)=& \exp(ik_\mathrm{L} \cos\theta) x+a\exp(-ik_\mathrm{L} \cos\theta) x,\\
  	\varphi_\mathrm{R}(x)=& b\exp(ik_\mathrm{R} \cos \phi) x,
\end{cases}
\end{equation}
where $a$ and $b$ denote the reflection and transmission coefficients, respectively.
The boundary conditions are described by
\begin{equation}
\begin{cases}
& \varphi_\mathrm{L}(0)= \varphi_\mathrm{R}(0),\\
&-\frac{\hbar^2}{2m_\mathrm{R}}\frac{\partial}{\partial x} \varphi_\mathrm{L}(x)|_{x=0}+\frac{\hbar^2}{2m_\mathrm{R}}\frac{\partial}{\partial x} \varphi_\mathrm{R}(x)|_{x=0}=H_\mathrm{b} \varphi_\mathrm{L}(0).
\end{cases}
\end{equation}
By applying the plane wave approximation, we obtain
($\frac{\hbar k_\mathrm{L}}{m_\mathrm{L}}\cos\theta \sim v_\mathrm{L} \cos\theta$,
$\frac{\hbar k_\mathrm{R}}{m_\mathrm{R}}\cos\phi \sim v_\mathrm{R}\cos\phi '$,
$v_\mathrm{L (R)}$ is the group velocity in the left (right) electrode),
\begin{align}
i\{\frac{\hbar v_\mathrm{R} b \cos\phi '}{2}-\frac{\hbar  v_\mathrm{L} (1-a) \cos\theta}{2}\}=bH_\mathrm{b}.
\end{align}
Note that $\phi$ is not equal to $\phi'$ because the directions of the wave vector and group velocity are not equivalent in the electron states with an anisotropic Fermi surface.
The transmission probability is given by
\begin{align}\label{sigmaNappB}
\sigmaN(\theta)=\frac{4 v_\mathrm{R} v_\mathrm{L}\cos\phi '\cos\theta}{(v_\mathrm{R}\cos\phi '+v_\mathrm{L}\cos\theta)^2+4\frac{H_\mathrm{b}^2}{\hbar^2}}.
\end{align}
We can verify that the flux conservation law for the $x$-axis direction
\begin{align}
(1-|a|^2)v_\mathrm{L}\cos\theta=|b|^2v_\mathrm{R}\cos\phi '
\end{align}
is satisfied in the present formulation.
\par
As an example, we consider the tunneling junction with an isotropic Fermi surface shape for the left electrode
($k_x^2+k_y^2=\frac{2 m E_\mathrm{F}}{\hbar^2}$) and an anisotropic shape for the right electrode,
$\left|k_x\right|^n+\left|k_y\right|^n=\frac{2.4mE_\mathrm{F}}{\hbar^2}$
with the tilting angle $\alpha$ (Fig.~4(b)).
The calculated results with $n=1.2$, $\alpha=0, \pi/8$, and $Z=5$ are shown in Fig.~4(c).
We can identify that the $\theta$-inversion symmetry in $\sigmaN(\theta)$ is lost in the $\alpha=\pi/8$ case as a result of broken $\theta$-inversion symmetry of the electronic states.
\par
Because the Fermi surface of real materials deviates from an isotropic shape, we expect that $\sigmaN(\theta)$ should be asymmetric in many realistic experimental situations.
Therefore, the assumption of asymmetric $\sigmaN(\theta)$ in the main text is reasonable.
However, as far as we know, no similar discussion has previously been presented.
The reason for this lack of discussion is that, for tunneling junctions with a TRS invariant superconductor, electron--hole symmetry $\sigmaT(E)=\sigmaT(-E)$ with tunneling spectra is applied even in cases of junctions with asymmetric $\sigmaN(\theta)$.
The symmetry of $\sigmaN(\theta)$ mentioned in this section does not need to be considered.
As claimed in this work, the broken electron--hole symmetry ($\sigmaT(E)\neq\sigmaT(-E)$) becomes apparent only in the conductance spectra for broken TRS superconductors.
In addition, checking the relation of $\sigmaT(E)=\sigmaT^R(-E)$ in the applied field is critically important for certifying the broken TRS because the origins of the broken electron--hole symmetry are not limited to the broken TRS of the superconductor.
\par
We note the important potential of tunneling spectroscopy based on the present analysis.
Tunneling spectroscopy has been accepted as preserving a higher energy resolution than other spectroscopies such as photoemission spectroscopy. 
However, its weakness is that it lacks momentum resolution and the results obtained are always integrals in momentum space. 
The results of the present study show that the weighting in momentum space can be controlled by tuning the relative Fermi surface of the two electrodes. 
This finding suggests the possibility of future major advances in tunneling spectroscopy with momentum space resolution.
\begin{figure*}
\includegraphics[keepaspectratio,width=1.8\columnwidth]{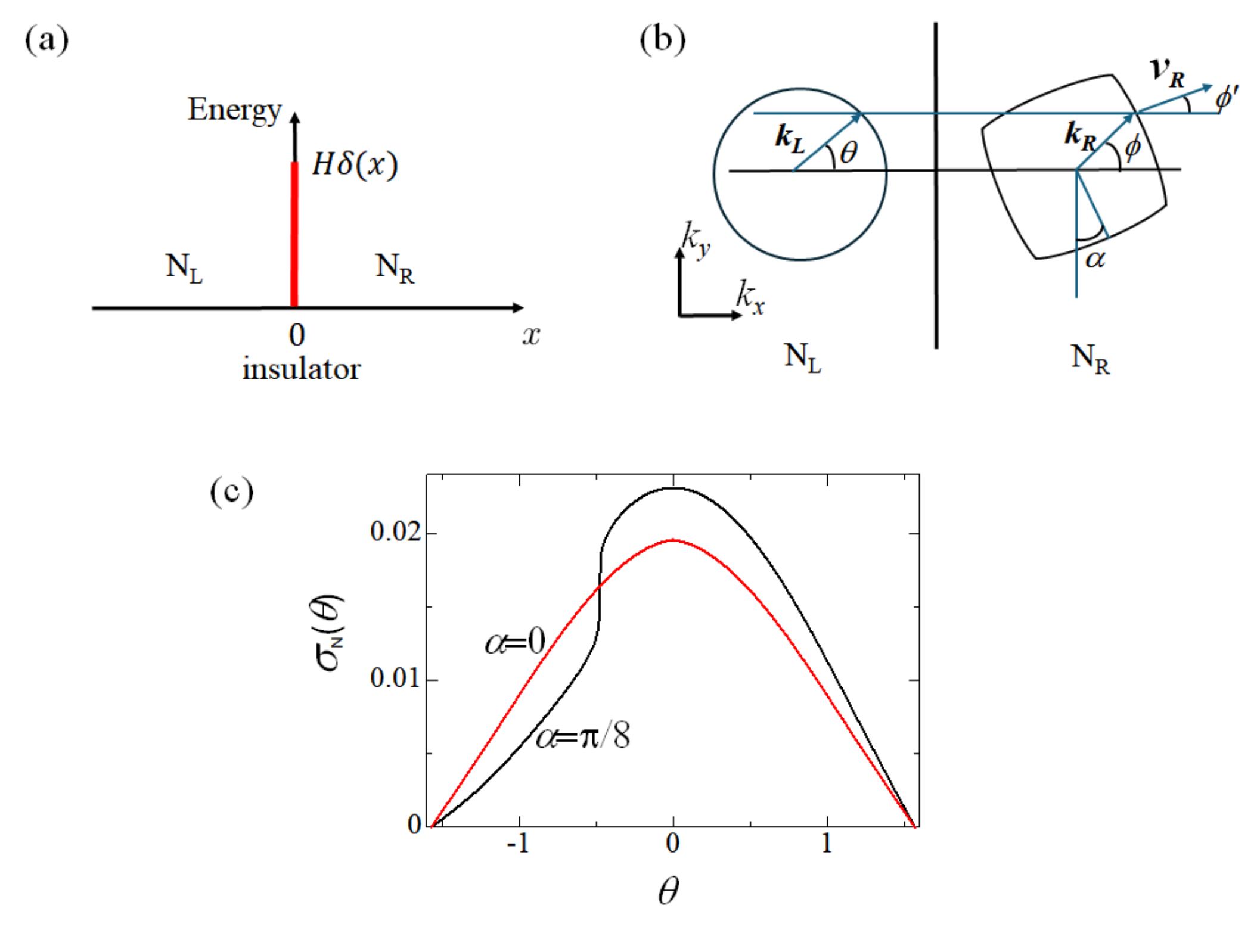}
	\caption{
	(a) Schematic of a tunneling junction with a $\delta$-function potential.
	(b) A tunneling junction model with broken inversion symmetry to $\theta$. An isotropic Fermi surface shape for the left electrode and an anisotropic shape with tilting angle $\alpha$ for the right electrode are assumed. The $y$ component of wavelength is conserved. The direction of the wave vector and group velocity are not equal in the right electrode.
	(c) Plots of calculated $\sigmaN(\theta )$ for $\alpha$=0 and $\pi$/8.}
	\label{fig2appB}
\end{figure*}

\section{B: Electron--hole symmetry in the extended BTK formula}

We here discuss the relation between the electron--hole symmetry, injection angle ($\theta$) symmetry, and time-reversal symmetry embedded in the extended BTK formula.
The purpose of the analysis is to verify the relations presented in Table I of the main text.
As a model, we consider a normal/insulator/superconductor (N/I/S) junction with equivalent isotropic Fermi surfaces in both electrodes.
The pair potential in the superconductor is generally represented by the mixture of spin-singlet (s) and spin-triplet (t) terms, and each term has even (e) and odd (o) components described by
\begin{align}
\Delta (\theta)=\Delta_\mathrm{se}(\theta)+\Delta_\mathrm{so}(\theta)+\Delta_\mathrm{te}(\theta)+\Delta_\mathrm{to}(\theta).
\end{align}
Each component satisfies the relation
\begin{align}
\begin{cases}
\Delta_\mathrm{se,so}(\theta)= \Delta_\mathrm{se,so}(\theta+\pi),\\
\Delta_\mathrm{te,to}(\theta)= -\Delta_\mathrm{te,to}(\theta+\pi),\\
\Delta_\mathrm{se,te}(\theta)= \Delta_\mathrm{se,te}(-\theta),\\
\Delta_\mathrm{so,to}(\theta)= -\Delta_\mathrm{so,to}(-\theta).\\
\end{cases}
\end{align}
\par
The extended BTK formula assumes the transmission process in which quasiparticles are injected with angle $\theta$ to the interface normal~\cite{BTK, PhysRevLett.74.3451, Kashiwaya1996,PhysRevB.56.7847, Kashiwaya2000}.
Four trajectories after the scattering are normal reflection, Andreev reflection, electron-like transmission, and hole-like transmission (Fig.~3(a)).
The conductance formula includes two effective pair potentials corresponding to the electron-like $\Delta_{+}(\theta)$ and hole-like transmission $\Delta_{-}(\theta)$.
Although the formula described in Eqs. (2)--(4) in the main text neglects the spin index because of the complexity, here we explicitly include it:
\begin{widetext}
\begin{align}
\sigma_{\mathrm{S}\uparrow(\downarrow)}(E,\theta) &= 
\frac{1+\sigmaN(\theta)|\Gamma_{+\uparrow(\downarrow)}(E,\theta)|^2
+(\sigmaN(\theta)-1)|\Gamma_{+\uparrow(\downarrow)}(E,\theta)\Gamma_{-\uparrow(\downarrow)}(E,\theta)|^2}
{|1+(\sigmaN(\theta)-1)\Gamma_{+\uparrow(\downarrow)}(E,\theta)\Gamma_{-\uparrow(\downarrow)}(E,\theta)|^2},\\
\Gamma_{+\uparrow(\downarrow)}(E,\theta)&=\frac{\Delta_{+\uparrow(\downarrow)}^{*}(\theta)}{E+\Omega_{+\uparrow(\downarrow)}(E,\theta)},
\Gamma_{-\uparrow(\downarrow)}(E,\theta)=\frac{\Delta_{-\uparrow(\downarrow)}(\theta)}{E+\Omega_{-\uparrow(\downarrow)}(E,\theta)},\\
\Omega_{\pm\uparrow(\downarrow)}(E,\theta)&=
\begin{cases}
\sqrt{E^2-|\Delta_{\pm\uparrow(\downarrow)}(\theta)|^2} & (E>|\Delta_{\pm\uparrow(\downarrow)}(\theta)|),\\
i\sqrt{|\Delta_{\pm\uparrow(\downarrow)}(\theta)|^2-E^2} &( -|\Delta_{\pm\uparrow(\downarrow)}(\theta)|\leq E\leq |\Delta_{\pm\uparrow(\downarrow)}(\theta)|),\\
-\sqrt{E^2-|\Delta_{\pm\uparrow(\downarrow)}(\theta)|^2} & (E<-|\Delta_{\pm\uparrow(\downarrow)}(\theta)|).
\end{cases}
\end{align}
\end{widetext}
The conductance is the function of the electron (or hole) and spins given by 
$\Delta_{+\uparrow}(\theta)$,
$\Delta_{-\uparrow}(\theta)$,
$\Delta_{+\downarrow}(\theta)$,
$\Delta_{-\downarrow}(\theta)$.
The total conductance spectrum is given by the integration of all injection angles,
\begin{align}
\sigmaT(E) =
\frac{\int_{-\pi/2}^{\pi/2}\left\{\sigma_{\mathrm{S}\uparrow}(E,\theta)+\sigma_{\mathrm{S}\downarrow}(E,\theta)\right\}
\sigmaN(\theta)\cos\theta \mathrm{d}\theta}
{\int_{-\pi/2}^{\pi/2}2\sigmaN(\theta)\cos\theta \mathrm{d}\theta}. 
\end{align}
The effective pair potentials are described by,
\begin{align}\label{eq:Delta(theta)}
\begin{split}
\Delta_{+\uparrow}(\theta ) &= \Delta_{\mathrm{se}}(\theta)
+\Delta_{\mathrm{so}}(\theta)
+\Delta_{\mathrm{te}}(\theta)
+\Delta_{\mathrm{to}}(\theta),\\
\Delta_{-\uparrow}(\theta ) &= \Delta_{\mathrm{se}}(\theta)
-\Delta_{\mathrm{so}}(\theta)
-\Delta_{\mathrm{te}}(\theta)
+\Delta_{\mathrm{to}}(\theta),\\
\Delta_{+\downarrow}(\theta ) &= -\Delta_{\mathrm{se}}(\theta)
-\Delta_{\mathrm{so}}(\theta)
+\Delta_{\mathrm{te}}(\theta)
+\Delta_{\mathrm{to}}(\theta),\\
\Delta_{-\downarrow}(\theta ) &= -\Delta_{\mathrm{se}}(\theta)
+\Delta_{\mathrm{so}}(\theta)
-\Delta_{\mathrm{te}}(\theta)
+\Delta_{\mathrm{to}}(\theta).
\end{split}
\end{align}
The effective pair potentials for the injection angle $-\theta$ satisfy
\begin{align}\label{eq:Delta(-theta)}
\begin{split}
\Delta_{+\uparrow}(-\theta ) &= \Delta_{\mathrm{se}}(\theta)
-\Delta_{\mathrm{so}}(\theta)
+\Delta_{\mathrm{te}}(\theta)
-\Delta_{\mathrm{to}}(\theta),\\
\Delta_{-\uparrow}(-\theta ) &= \Delta_{\mathrm{se}}(\theta)
+\Delta_{\mathrm{so}}(\theta)
-\Delta_{\mathrm{te}}(\theta)
-\Delta_{\mathrm{to}}(\theta),\\
\Delta_{+\downarrow}(-\theta ) &= -\Delta_{\mathrm{se}}(\theta)
+\Delta_{\mathrm{so}}(\theta)
+\Delta_{\mathrm{te}}(\theta)
-\Delta_{\mathrm{to}}(\theta),\\
\Delta_{-\downarrow}(-\theta ) &= -\Delta_{\mathrm{se}}(\theta)
-\Delta_{\mathrm{so}}(\theta)
-\Delta_{\mathrm{te}}(\theta)
-\Delta_{\mathrm{to}}(\theta).
\end{split}
\end{align}
We find the relation in the effective pair potentials given by
\begin{align}\label{eq:Delta}
\begin{split}
\Delta_{+\uparrow(\downarrow)}(\theta ) &= -\Delta_{-\downarrow(\uparrow)}(-\theta),\\
\Delta_{-\uparrow(\downarrow)}(\theta ) &= -\Delta_{+\downarrow(\uparrow)}(-\theta).
\end{split}
\end{align}
Through the conversion $E \to -E$ and $ \theta \to -\theta$, $\Omega_{\pm \uparrow(\downarrow)}(E,\theta)$ and $\Gamma_{\pm \uparrow(\downarrow)}(E,\theta)$ have the relations described by
\begin{align}\label{eq:Omega}
\begin{split}
\Omega_{+\uparrow(\downarrow)}(E,\theta) &= -\Omega_{-\downarrow(\uparrow)}^{*}(-E,-\theta),\\
\Omega_{-\uparrow(\downarrow)}(E,\theta) &= -\Omega_{+\downarrow(\uparrow)}^{*}(-E,-\theta),\\
\Gamma_{+\uparrow(\downarrow)}(E,\theta) &= -\Gamma_{-\downarrow(\uparrow)}^{*}(-E,-\theta),\\
\Gamma_{-\uparrow(\downarrow)}(E,\theta) &= -\Gamma_{+\downarrow(\uparrow)}^{*}(-E,-\theta).
\end{split}
\end{align}
\par
We first consider cases i) and ii) in Table I when $\sigmaN(\theta)$ is $\theta$ symmetric.
The conductance $\sigma_{\mathrm{S}\uparrow(\downarrow)}(E,\theta)$ is described by 
\begin{widetext}
\begin{align}
\begin{split}
\sigma_{\mathrm{S}\uparrow(\downarrow)}(E,\theta)
&= \frac{1+\sigmaN(\theta)|\Gamma_{+\uparrow(\downarrow)}(E,\theta)|^2
+(\sigmaN(\theta)-1)|\Gamma_{+\uparrow(\downarrow)}(E,\theta)\Gamma_{-\uparrow(\downarrow)}(E,\theta)|^2}
{|1+(\sigmaN(\theta)-1)\Gamma_{+\uparrow(\downarrow)}(E,\theta)\Gamma_{-\uparrow(\downarrow)}(E,\theta)|^2}\\
&= \frac{1+\sigmaN(-\theta)|\Gamma_{-\downarrow(\uparrow)}^{*}(-E,-\theta)|^2
+(\sigmaN(-\theta)-1)|\Gamma_{-\downarrow(\uparrow)}^{*}(-E,-\theta)\Gamma_{+\downarrow(\uparrow)}^{*}(-E,-\theta)|^2}
{|1+(\sigmaN(-\theta)-1)\Gamma_{-\downarrow(\uparrow)}^{*}(-E,-\theta)\Gamma_{+\downarrow(\uparrow)}^{*}(-E,-\theta)|^2}\\
&= \sigma_{\mathrm{S}\downarrow(\uparrow)}(-E,-\theta).
\end{split}
\end{align}
Thus, the total conductance $\sigmaT(E)$ is electron--hole symmetric,
\begin{align}
\begin{split}
\sigmaT(E) &=
\frac{\int_{-\pi/2}^{\pi/2}\left\{\sigma_{\mathrm{S}\uparrow}(E,\theta)+\sigma_{\mathrm{S}\downarrow}(E,\theta)\right\}
\sigmaN(\theta)\cos\theta \mathrm{d}\theta}
{\int_{-\pi/2}^{\pi/2}2\sigmaN(\theta)\cos\theta \mathrm{d}\theta}\\
&=
\frac{\int_{-\pi/2}^{\pi/2}\left\{\sigma_{\mathrm{S}\uparrow}(-E,-\theta)+\sigma_{\mathrm{S}\downarrow}(-E,-\theta)\right\}
\sigmaN(-\theta)\cos\theta \mathrm{d}\theta}
{\int_{-\pi/2}^{\pi/2}2\sigmaN(-\theta)\cos\theta \mathrm{d}\theta}\\
&=\sigmaT(-E).
\end{split}
\end{align}
\end{widetext}
\par
Next, we consider case iii) in Table I when $\sigma_N(\theta)$ is $\theta$-asymmetric and the superconductor maintains the TRS ($\Delta_{\pm\uparrow\downarrow}(\theta) = \Delta_{\pm\uparrow\downarrow}^{*}(\theta)$).
In this case, we obtain the relations
\begin{align}\label{eq:Gamma-TRS}
\begin{split}
\Omega_{+\uparrow(\downarrow)}(E,\theta) &= -\Omega_{+\uparrow(\downarrow)}^{*}(-E,\theta),\\
\Omega_{-\uparrow(\downarrow)}(E,\theta) &= -\Omega_{-\uparrow(\downarrow)}^{*}(-E,\theta),\\
\Gamma_{+\uparrow(\downarrow)}(E,\theta) &= -\Gamma_{+\uparrow(\downarrow)}^{*}(-E,\theta),\\
\Gamma_{-\uparrow(\downarrow)}(E,\theta) &= -\Gamma_{-\uparrow(\downarrow)}^{*}(-E,\theta).
\end{split}
\end{align}
By applying these relations to $\sigma_{\mathrm{S}\uparrow(\downarrow)}(E,\theta)$, we obtain 
\begin{align}\label{eq:sigmaS-TRS}
\sigma_{S\uparrow(\downarrow)}(E,\theta)= \sigma_{S\uparrow(\downarrow)}(-E,\theta).
\end{align}
On the basis of this relation, we obtain the electron--hole symmetry of the BTK formula written by $\sigmaT(E) = \sigmaT(-E)$.
\par
Finally, we consider case iv) in Table I, where $\sigmaN(\theta)$ is $\theta$-asymmetric and the superconductor breaks TRS ($\Delta_{\pm\uparrow\downarrow}(\theta) \neq \Delta_{\pm\uparrow\downarrow}^{*}(\theta)$).
In this case, we are unable to obtain the electron--hole symmetry of the BTK formula.
However, we obtain a new relation by considering the conductance for the superconductor with time-reversal (conjugate) pair potential $\Delta^*(\theta)$.
We use superscript ``C'' to describe the quantities for the conjugate pair potential.
\begin{align}\label{eq:Gamma-TRS2}
\begin{split}
\Omega_{+\uparrow(\downarrow)}(E,\theta)&=\Omega_{+\uparrow(\downarrow)}^\mathrm{C}(-E,\theta),\\
\Omega_{-\uparrow(\downarrow)}(E,\theta)&=\Omega_{-\uparrow(\downarrow)}^ \mathrm{C}(-E,\theta),\\
\Gamma_{+\uparrow(\downarrow)}^{*}(E,\theta)&=-\Gamma_{+\uparrow(\downarrow)}^ \mathrm{C}(-E,\theta),\\
\Gamma_{-\uparrow(\downarrow)}^{*}(E,\theta)&=-\Gamma_{-\uparrow(\downarrow)}^ \mathrm{C}(-E,\theta).
\end{split}
\end{align}
By applying these relations to $\sigma_{\mathrm{S}\uparrow(\downarrow)}(E,\theta)$, we obtain
\begin{align}\label{eq:sigmaS-TRSB2}
\sigma_{S\uparrow(\downarrow)}(E,\theta)= \sigma_{S\uparrow(\downarrow)}^C(-E,\theta),
\end{align}
and the total conductance has the relation $\sigmaT(E) = \sigmaT^C(-E)$.



\bibliography{reference_prx.bib}

\end{document}